\documentclass[12pt,preprint]{aastex}

\begin{document}


\title{A {\it Spitzer} Spectrum of the Exoplanet HD 189733b}

\author{C. J. Grillmair}
\affil{Spitzer Science Center, 1200 E. California Blvd., Pasadena,  CA 91125}
\email{carl@ipac.caltech.edu}
\author{D. Charbonneau\altaffilmark{1}}
\affil{Harvard-Smithsonian Center for Astrophysics, 60 Garden St.,
  MS-16, Cambridge, MA 02138}
\email{dcharbon@cfa.harvard.edu}
\author{A. Burrows}
\affil{Steward Observatory, University of Arizona, Tucson, AZ 85721}
\email{burrows@.as.arizona.edu}
\author{L. Armus}
\affil{Spitzer Science Center, 1200 E. California Blvd., Pasadena,  CA 91125}
\email{lee@ipac.caltech.edu}
\author{J. Stauffer}
\affil{Spitzer Science Center, 1200 E. California Blvd., Pasadena,  CA 91125}
\email{stauffer@ipac.caltech.edu}
\author{V. Meadows}
\affil{Spitzer Science Center, 1200 E. California Blvd., Pasadena,  CA 91125}
\email{vsm@ipac.caltech.edu}
\author{J. Van Cleve}
\affil{Ball Aerospace \& Technologies Corp., PO Box 1062, Boulder, CO 80306}
\email{jvanclev@ball.com}
\and
\author{D. Levine}
\affil{Spitzer Science Center, 1200 E. California Blvd., Pasadena,  CA 91125}
\email{deblev@ipac.caltech.edu}

\altaffiltext{1}{Alfred P. Sloan Research Fellow}

\begin{abstract}

We report on the measurement of the 7.5-14.7 micron spectrum for the
transiting extrasolar giant planet HD 189733b using the Infrared
Spectrograph on the Spitzer Space Telescope. Though the observations
comprise only 12 hours of telescope time, the continuum is well
measured and has a flux ranging from 0.6 mJy to 1.8 mJy over the
wavelength range, or $0.49 \pm 0.02\%$ of the flux of the parent star.
The variation in the measured fractional flux is very nearly flat over
the entire wavelength range and shows no indication of significant
absorption by water or methane, in contrast with the predictions of
most atmospheric models.  Models with strong day/night differences appear
to be disfavored by the data, suggesting that heat redistribution to
the night side of the planet is highly efficient.

\end{abstract}



\section{Introduction}

HD 189733 is a nearby K0V star with a magnitude of K = 5.5 and an
expected flux of 250 mJy at 10 microns.  HD 189733b was initially
discovered in the ELODIE search for transiting exoplanets
\citep{bouchy2005} and subsequently detected in existing Hipparcos
data \citep{hebrard2006}. In addition to HD 189733b, HD 189733 is
orbited by an M dwarf at a distance of 216 AU \citep{bakos2006a}. HD
189733b orbits its parent star in 2.2 days and has an estimated radius
of $R_p = 1.154 \pm 0.033 R_J$ \citep{bakos2006b}. From first to last
contact, the duration of the secondary eclipse is $\approx 1.9$ hours.

The planet-to-star surface area ratio for HD 189733b is $\approx
0.024$ \citep{bakos2006b}.  Owing to an orbital radius of 0.031 AU,
the upper atmospheric temperature of the planet is $\geq 1000$ K
\citep{fortney2006a}. Together these factors yield an expected
planet-to-star flux ratio that is almost a factor of two greater at 10
microns than for either TrES-1 or HD 209458b, two other exoplanets
whose thermal emission has recently been detected by the Spitzer Space
Telescope \citep{charbonneau2005, deming2005}.  This expectation has
been borne out by \citet{deming2006}, who used the Peak-up Imaging
array portion of {\it Spitzer's} Infrared Spectrograph (IRS) to
measure a secondary eclipse depth at $16 \micron$ of $0.551 \pm
0.030\%$.  This makes HD 189733b a prime candidate for IRS
spectroscopy and a more detailed characterization of its physical
properties.

We briefly describe the observations in Section \ref{observations}
and the analysis in Section \ref{analysis}. We discuss the
spectrum of HD 189733b in Section \ref{discussion}.

\section{Observations \label{observations}}

HD 189733 was observed with the Infrared Spectrograph (IRS) on October
21st and November 21st, 2006 as part of General Observer program
\#30473. In each case the star was observed for a six-hour span
centered on the predicted time of secondary eclipse.  All observations
were made in first order ($\lambda = 7.4-14.5 \micron$) using the
Short-Low module, giving a spectral resolution of $\approx 80$.  The
performance and capabilities of the Short-Low module, which has the
cleanest and most sensitive of the IRS detectors, is discussed in
detail by \citet{houck2004}.

The Oct 21st observations were carried out in ``Staring'' mode,
wherein the telescope was periodically repointed to place the target
alternately at two dither positions along the slit.  Alternating
between two slit positions provided some insurance against the
possible large buildup of latent charge (see below). The observations
were preceded by a high-accuracy peak-up approximately 3 hours before
secondary eclipse and were taken continuously through the
out-of-eclipse, ingress, secondary eclipse, egress, and out-of-eclipse
portions of HD 189733b's orbit. Thirty successive 14-second exposures
were taken at each slit position before moving to the next dither
position. A total of 30 telescope moves were commanded, yielding 450
independent integrations at each dither position over the course of
six hours.

A preliminary analysis of the Oct 21st observations showed that charge
latents were considerably more benign than initial estimates.  An
observing sequence scheduled for Nov 21st was consequently modified to
place the target at the center of the Short-Low slit for the entire
observing sequence. Small pointing errors associated with any
commanded telescope movement (which contribute significantly to the
noise budget) were thus avoided. Having thereby also eliminated
telescope slew and settle times, we were able to fit a total 950
14-second exposures into the six hours available.

\section{Analysis \label{analysis}}

Our analysis makes use of the two-dimensional Basic Calibrated Data
products from the S14 version of the IRS pipeline. Using Version 1.4
of the SPICE spectral extraction tool, we apply the optimal
extraction feature to extract one-dimensional spectra for each of the
1850 exposures separately. We fix the center of each extraction window
to take advantage of the pointing performance of the telescope and
avoid adding noise associated with a measurement of the
sub-pixel position of the star for each exposure.

For reasons that will become clear, we treat the integrations at
the two dither positions on Oct 21st and the single-pointing
integrations taken on Nov 21st as three separate sets of
observations. In principal, it should be possible to simply coadd all
integrations in each observation set taken during eclipse (starlight
only) and subtract them from a coaddition of all spectra taken out of
eclipse (star plus planet). In practice, due to the limited
integration time available when the planet is in eclipse, several
systematics can manifest themselves as either increased noise or
spectral offsets.

The Spitzer Space Telescope's stationary pointing performance is not
perfect. Due to a small, cyclic heat source within the spacecraft
structure, there is a periodic variation in alignment between the Star
Tracker assembly and the telescope boresight. This leads to a pointing
oscillation with a period of about an hour and a magnitude of $\sim
0.05\arcsec$ \citep{morales2006}. Though small, this oscillation has
significant consequences for observations that require precision on
the order of 0.1\%. In addition to variations in the total
amount of light entering the slit, we are subject to significant signal
modulation due to drifts across pixels with different (and poorly
calibrated) sensitivities.

We analyze each wavelength bin as an independent time series and then
combine the results to produce a final ratio spectrum. As an example,
we show in Figure 1 the measured flux as a function of time at dither
position 1 for the wavelength bin centered on $7.93 \micron$ . Several
characteristics that affect all wavelength bins to a greater or
lesser degree are in evidence: 1) There is a rapid rise in the
counts during the first 10 minutes of the sequence that is clearly
related to latent charge buildup. 2) There is a longer term variation
that is probably related to both the buildup and decay of latent
charge, as well as to possible drifts in and out of the slit. 3) There
is a cyclic behavior due to the pointing oscillations.

We have found that the buildup and decay of latent charge can vary
greatly from one pixel to the next. Rather than trying to model the
ensemble, short-term latent behavior for each wavelength bin, we
simply exclude the first 10 minutes of data from our analysis. Since
our final signal-to-noise ratio is determined primarily by
data taken during secondary eclipse, the loss of the first 10 minutes
of data has negligible consequences for the final result.  We fit the
longer-term variation with a 3rd-order polynomial and then divide it
out to flatten the time series. We find this to be the minimum order
necessary to reproduce the observed drifts while leaving the periodic
variations largely unaffected.

Observations with the InfraRed Array Camera (IRAC) indicate that the
Spitzer Space Telescope's pointing oscillation behaves rather more
like a sawtooth than a sinusoid \citep{morales2006}.  We indeed find
that an asymmetric sawtooth best fits the cyclic time behavior in the
data, and that the shape of the sawtooth is fairly consistent from one
slit position to the next. We find that the phase and asymmetry of the
flux modulation (which depend on both the telescope motion and the
distribution of imperfectly calibrated pixels) are essentially
constant for all wavelength bins (modulo a $180\arcdeg$ phase flip),
and we adopt a fixed modulation pattern, with only the amplitude of
the oscillation being a free parameter. The sawtooth that best fits
the data for dither position 1 is shown over plotted on the
drift-corrected data in the middle panel of Figure 1.

After dividing the data by the best fitting periodic function, we fit
the data using the ephemerides and the secondary eclipse profile of
\citet{bakos2006b}. The uncertainty in the predicted time of the
center of the secondary eclipse is 5.4 minutes assuming a circular
orbit. The ephemerides we have adopted are consistent with updated
values recently published by \citet{winn2007}. The timing
uncertainty has been reduced to 0.5 minutes and is not a significant source
of uncertainty in our analysis. In fitting the light curve, the two
free parameters are the mean flux level outside of eclipse and the
scaling of Bakos et al.'s light curve required to match the in-eclipse
data. This allows us to use data taken during ingress and egress as
well as during totality. We optimize the fit by varying the amplitude
of the periodic function and minimizing $\chi^2$ during the in-eclipse
portion of HD 189733b's orbit. Due to its greater extent, the mean
level of the time series outside of eclipse is largely unaffected by
changes in the periodic function, and can be independently measured to
high precision.

To constrain the non-periodic behavior of pixels in each wavelength
bin over the course of six hours, we first divide the pre-correction,
in-eclipse data with the light curve of \citet{bakos2006b}.  We
find that simply fitting to all data with $t < -2500$ s and $t > 2500$
s is not sufficient to constrain long term drifts during eclipse and
adds noticeably to the noise in the final spectrum.  Bakos et al.'s
(2006) light curve is initially scaled so that the upward correction
at eclipse center is 0.5\%.  We then iterate on the final solution for
each wavelength bin by rescaling the correction based on the previous
flux estimate. The solution converges rapidly in all cases, and we
halt the calculation when the difference in computed eclipse depth
between successive iterations becomes less than 0.002\%.  Changing the
initial scaling of the light curve by a factor of two up or down has
no effect on the final spectrum.

Given a perfectly stable detector and platform, we would normally
expect to be dominated by photon noise over the entire wavelength
range for a star as bright as HD 189733.  After making the corrections
described above, the RMS spread of the individual measurements about
the best fit light curve are indeed within 20\% of the expected photon
noise limit over most of the wavelength range. The RMS spread in the
individual measurements exceeds what one would expect just from photon
noise by up to 50\% in the regions $9-10 \micron, 11.5-12.5 \micron$,
and $13.7-14.5 \micron$ for both the Oct 21st and Nov 21st observations.
Conversely, the regions $8-9 \micron, 10-11 \micron$, and $12.5-13.5
\micron$ show an RMS spread that is consistent with pure photon
statistics. This suggests that there are remaining, unmodeled temporal
variations on the order of 0.5\%, due perhaps to fringe motions and
subpixel variations in responsivity.

In Figure 2 we show the HD 189733b/HD 189733 flux ratios for each of
the three separately analyzed data sets.  The plotted uncertainties
reflect only the light curve fitting errors and are smallest for the
Nov 21st observations by virtue of their greater number and continuous
coverage through secondary eclipse.

\section{Discussion \label{discussion}}

HD 189733b's continuum is well measured and, multiplying the relative
fluxes in Figure 2 by the observed spectrum of HD 189733, we obtain
absolute fluxes ranging from 1.8 mJy at $7.8 \micron$ to 0.5 mJy at
$14.7 \micron$. Our measurement is consistent with the 660 $\mu$Jy
integrated flux measurement at $16 \micron$ by \citet{deming2006}.
Comparing the three spectra to one another, the signal-to-noise ratio
per resolution element ranges from about 15 at 10 microns to about 6
at 14.5 microns.

The flux ratios in Figure 2 are essentially flat; the emission
spectrum of HD 189733b over the wavelength range 7.5 to $14.7 \micron$
is evidently consistent with that of a blackbody. The integrated flux
ratios are 0.50\%, 0.51\%, and 0.46\%, respectively, yielding a mean
of $0.49 \pm 0.02\%$. Though the mean relative flux is in good agreement,
the spectrum taken at dither position 1 appears somewhat problematic,
with a number of peaks and troughs that are not seen at the other
slit positions. Of the three, the spectrum taken at slit center is
clearly the highest quality measurement and demonstrates the utility
of keeping the telescope as motionless as possible for observations of
this type. While there are interesting similarities between the
spectra taken at dither position 2 and at slit center, the differences
are sufficiently large and numerous that we cannot identify individual
absorption or emission features with any confidence.

In Figure 3 we compare the relative flux spectrum measured at slit
center with a model of HD 189733b from \citet{burrows2006}. The model
shown is for superior conjunction, with no clouds and 50\% energy
redistribution to the night side of the planet.  The mean levels for
the model and the observations are quite similar. Redward of $10
\micron$, where the flux ratio is predicted to be fairly flat, the
model predicts a mean relative flux which is about 20\% higher than
that of the observations.  Near superior conjunction, less atmospheric
energy redistribution would push the relative planet/star flux still
higher.  In contrast to the findings of \citet{harrington2006} for
$\upsilon$ And b, it appears that strong day/night differences on HD
189733b are disfavored by the data.

Blueward of 8.2 microns, the model predicts a quite significant drop
in the relative flux due to water absorption. In contrast, the
observed flux ratios continue unabated to the blue end of the
spectrum. This difference is significant; scaling
\citet{burrows2006}'s model to match the data beyond $9.5 \micron$ and
summing over all bins blueward of 8.2 microns, the model lies $\approx
9\sigma$ below the data. The measured flux at the blue end is not the
product of a few bad measurements, but is reflected separately in each
of the three individual spectra, and two independent analyses. It is
also seen in simple subtractions of combined out-of-eclipse and
in-eclipse spectra.

A reduced planet-to-star ratio blueward of $10\micron$ due to water
opacity is a near universal prediction in published models of hot
Jupiters to date (e.g. \citet{fortney2005, barman2005, seager2005,
burrows2006}), and hence the absence of this signature from our data
is intriguing. Simply reducing the amount of water in the models may
not, by itself, bring them into agreement with the data.  For example,
our observations also appear to disagree with the low H$_2$O, high
carbon to oxygen ratio spectrum computed for HD 209458b by
\citet{seager2005}. In this case a similar drop in the flux blueward
of 9 microns is expected due to methane absorption. We note that
\citet{fortney2006b} have recently considered dynamical models of hot
Jupiter atmospheres \citep{cooper2006} to predict emergent spectra as
a function of orbital phase, and have shown that spectral features in
day-side spectra may be suppressed due to an isothermal
pressure-temperature profile. A more thorough analysis of the
consequences of this spectrum for models of giant planet atmospheres
is forthcoming.

The detection of HD 189733b's spectral continuum, accomplished with
only 12 hours of Spitzer time, is a remarkable demonstration of the
capability and utility of the Spitzer Space Telescope and the Infrared
Spectrograph. Integrating over many eclipses, we can expect
substantial increases in spectral signal-to-noise ratio, broader
wavelength coverage, and the detection of atmospheric constituents for
both HD 189733b and other transiting extrasolar giant planets. While
Spitzer's cryogen lasts, we can look forward to an exciting era of
quantitative, spectral characterization of extrasolar planets, and
ultimately a deeper understanding of planetary atmospheres under
strong irradiation.

\acknowledgments

We are grateful to an anonymous referee for several suggestions which
significantly improved the analysis and presentation of this
work. This work is based on observations made with the {\it Spitzer
Space Telescope}, which is operated by the Jet Propulsion Laboratory,
California Institute of Technology under a contract with NASA. Support
for this work was provided by NASA through an award issued by
JPL/Caltech. This study was supported in part by NASA grant NNGO4GL22G
and through the NASA Astrobiology Institute under Cooperative
Agreement No. CAN-02-OSS-02 issued through the Office of Space
Science.

\clearpage

\begin{figure}
\epsscale{0.8}
\plotone{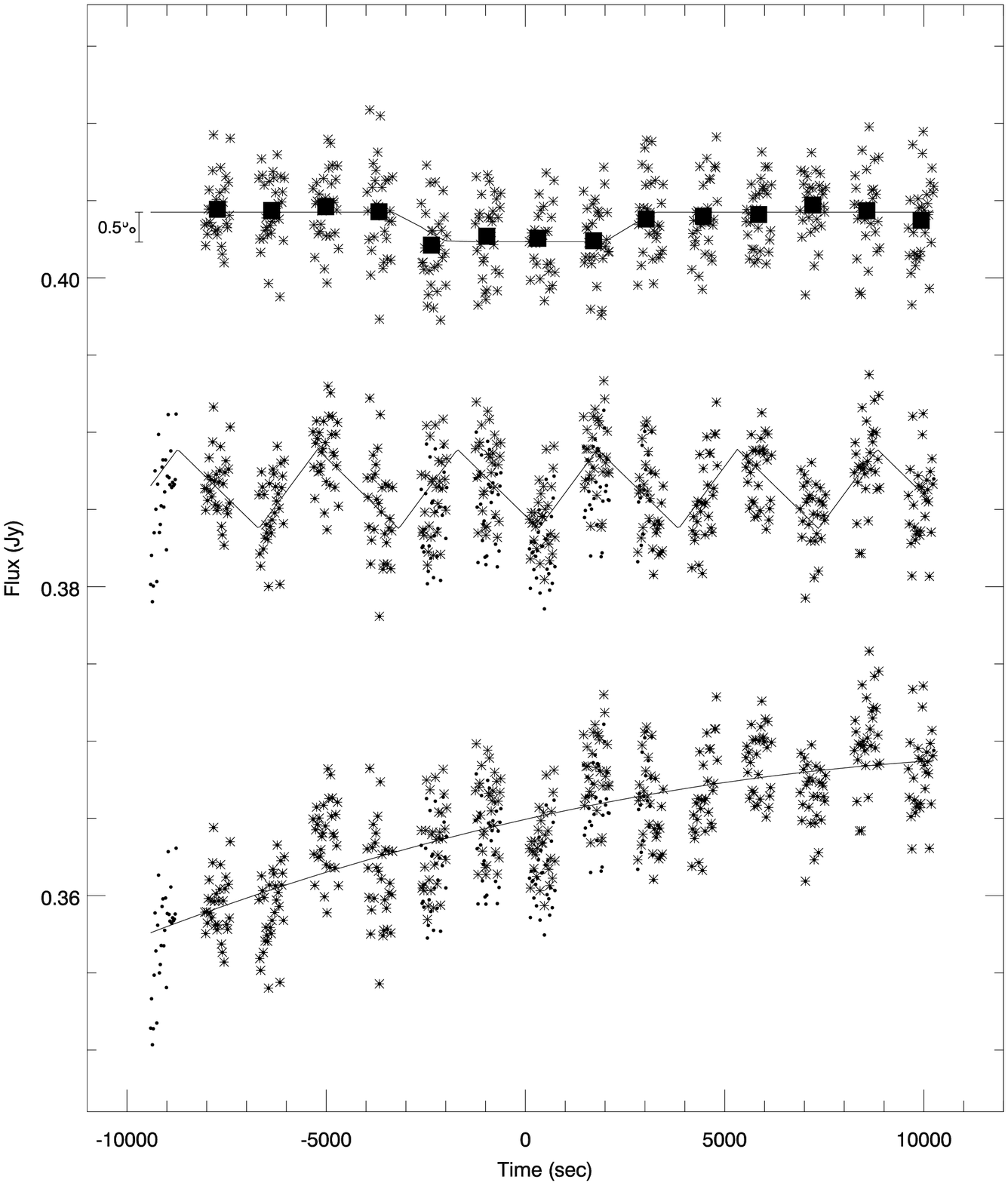}
\caption{The best-fit solution for the planetary flux in the
  wavelength bin centered on 7.93 microns, showing from bottom to top
  the corrections applied. The flux measurements are shown as a function
  of time from secondary eclipse. The upper and lower data
  sets have been vertically offset for clarity. The filled circles in
  the bottom plot show the uncorrected flux as extracted using
  SPICE. The asterisks highlight data points actually used to (1)
  constrain the long term drift, (2) fit the amplitude of the periodic
  function, and (3) fit the eclipse depth. To better constrain (1) and
  (2), the asterisked data have been corrected upwards during
  secondary eclipse using the predicted light curve of
  \citet{bakos2006b}. The first 10 minutes of data are not used as
  they are generally affected by rapid latent charge buildup. The
  middle set of data points shows the effect of dividing by a
  3rd-order polynomial, and over plotted is the sawtooth function that
  best fits the pointing-induced oscillations. The upper set of data
  points have been corrected for long term and periodic behavior, and
  the best fitting light curve is shown over plotted. The filled
  squares are each a mean of 30 successive data points and serve
  simply as an aid in visualizing the quality of the
  fit. \label{fig1}}
\end{figure}

\clearpage

\begin{figure}
\epsscale{1.0}
\plotone{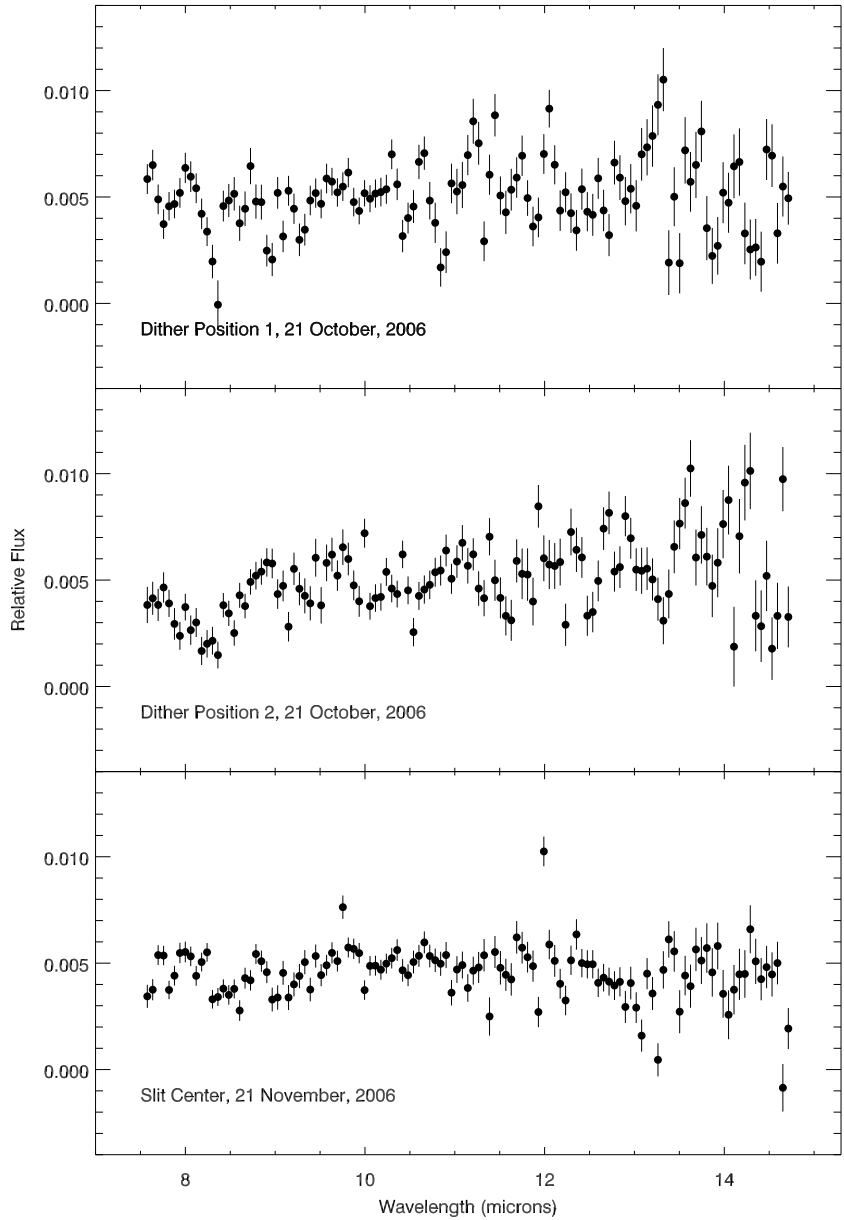}
\caption{The flux of HD 189733b relative to that of HD 189733, as
measured at three different slit positions.  The spectra are
unsmoothed, and the plotted uncertainties correspond to the formal
fitting errors.}
\end{figure}

\clearpage

\begin{figure}

\plotone{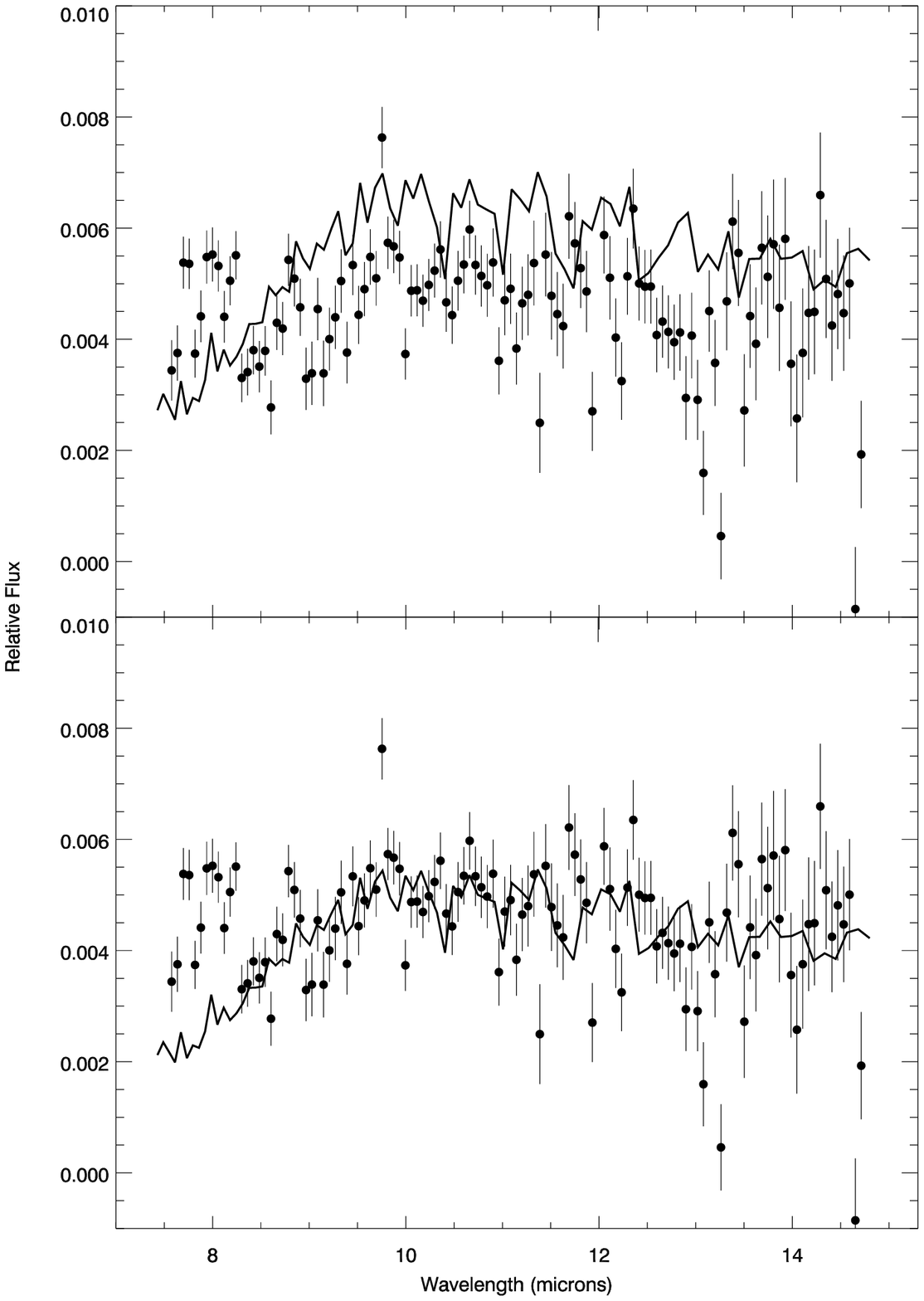}
\caption{Comparison of the flux ratios measured at slit center (filled
circles) with a model of HD 189733b from \citet{burrows2006}. The
model shown is for superior conjunction, with no clouds and 50\%
energy redistribution to the night side of the planet. The upper panel
shows the model as published by \citet{burrows2006}, while in the
lower panel the model has been scaled to match the data at $\lambda >
9.5 \micron$. Intriguingly, the observed spectrum does not show the
expected decrease in relative flux at short wavelengths due to
increasing water opacity.}
\end{figure}

\end{document}